\documentclass[a4paper]{article}

\usepackage{INTERSPEECH2022}
\usepackage{amsmath,graphicx}
\usepackage{pifont}
\usepackage{xcolor}
\usepackage[T1]{fontenc}
\usepackage[utf8]{inputenc}
\usepackage{babel}
\usepackage[font=small,labelfont=bf]{caption}

\newcommand{\cmark}{\ding{51}}%
\newcommand{\xmark}{\ding{55}}%

\title{Deploying self-supervised learning in the wild for hybrid automatic speech recognition}
\name{Mostafa Karimi, Changliang Liu, Kenichi Kumatani, Yao Qian, Tianyu Wu, Jian Wu}
\address{Microsoft Corporation, Redmond, WA}
\email{\{mkarimi, chanliu, kekumata, yaoqian, tianyu.wu, jianwu\}@microsoft.com}

\begin{document}

\maketitle
\begin{abstract}
Self-supervised learning (SSL) methods have proven to be very successful in automatic speech recognition (ASR)~\cite{baevski2020wav2vec,hsu2021hubert,chung2019unsupervised}.
These great improvements have been reported mostly based on highly curated datasets such as LibriSpeech~\cite{panayotov2015librispeech} for non-streaming End-to-End ASR models. However, the pivotal characteristics of SSL is to be utilized for any untranscribed audio data. In this paper, we provide a full exploration on how to utilize uncurated audio data in SSL from data pre-processing to deploying an streaming hybrid ASR model. More specifically, we present (1) the effect of Audio Event Detection (AED) model in data pre-processing pipeline (2) analysis on choosing optimizer and learning rate scheduling (3) comparison of recently developed contrastive losses, (4) comparison of various pre-training strategies such as utilization of in-domain versus out-domain pre-training data, monolingual versus multilingual pre-training data, multi-head multilingual SSL versus single-head multilingual SSL and supervised pre-training versus SSL. The experimental results show that SSL pre-training with in-domain uncurated data can achieve better performance in comparison to all the alternative out-domain pre-training strategies.
\end{abstract}

\section{Introduction}
\label{sec:intro}
Automatic Speech Recognition (ASR) typically requires a huge amount of transcribed audio data for reasonable performance~\cite{schneider2019wav2vec}. 
Expanding ASR to new languages with low resources of transcribed audio data still remains a challenge. To circumvent this issue, at first supervised cross-lingual pre-training from high resource languages to low resource language has been introduced~\cite{kunze2017transfer,huang2013cross}. Supervised cross-lingual pre-training is categorized to monolingual transfer~\cite{kunze2017transfer} or multilingual transfer learning by sharing the hidden layers across languages and considering language-dependent softmax layers~\cite{huang2013cross}. However, since transcribed audio data are scarce, very expensive and time-consuming to obtain, industry and academia has shown great interest in unsupervised pre-training methods.

Unsupervised pre-training mostly known as Self-supervised learning (SSL) has shown outstanding performances in various domains such as Computer Vision (CV)~\cite{chen2020simple,goyal2021self}, Natural Language Processing (NLP)~\cite{devlin2018bert} and ASR~\cite{baevski2020wav2vec,hsu2021hubert,chung2019unsupervised,zhang2020pushing,hsu2021robust,sadhu2021wav2vec}. Most of the research has been focused on in-domain monolingual SSL methods such as wav2vec 2.0~\cite{baevski2020wav2vec}, HuBERT~\cite{hsu2021hubert}, and APC~\cite{chung2019unsupervised}. These SSL models have been developed and trained on well-known, human-curated and annotated datasets such as LibriSpeech~\cite{panayotov2015librispeech} and LibriVox. Recently, SUPERB~\cite{yang2021superb} has provided a thorough benchmark for various downstream tasks such as ASR, speaker identification, keyword spotting and etc. SUPERB furthermore demonstrates the effectiveness of SSL in the speech domain. In addition, a multilingual unsupervised pre-training strategy also has gained interest through cross-lingual representation learning~\cite{conneau2020unsupervised,wang2021unispeech}. However again they have been trained and developed through well-curated datasets and mostly on non-streaming End-to-End scenarios. The great premise of SSL is to be utilized for any accessible audio data such as news, podcasts, interviews for streaming ASR which is nowadays used in most of production settings. Therefore the main question is "how to develop a fully automated unsupervised pre-training pipeline for ASR based on wild audio data?". 

In this paper, we present and explore the whole pipeline of SSL on streaming hybrid ASR from data pre-processing, choosing training recipes and optimizer, comparing contrastive loss functions. Moreover, we thoroughly compare various pre-training strategies. To be more specific, the major contributions of our paper are (i) \emph{data pre-processing}: We explore the effectiveness of Xception-based~\cite{chollet2017xception} Audio event detection (AED) model on pre-processing pipeline; (ii) \emph{Training}: we demonstrate the effect of optimizer and learning rate scheduling in SSL for streaming hybrid ASR; (iii) \emph{Loss function}: We utilize the recently developed contrastive loss function in computer vision such as flatNCE~\cite{chen2021simpler} to overcome drawbacks of infoNCE~\cite{baevski2020wav2vec}, namely large bias in estimating mutual information and sensitivity to minibatch size; (iv) \emph{Analysis}: we thoroughly compare pre-training strategies by considering all the aspects such as monolingual versus multilingual pre-training, out-domain versus in-domain pre-training, single-head versus multi-head multilingual SSL and supervised versus unsupervised pre-training.

\section{RELATION TO PRIOR WORK}
\label{sec:prior}
SSL in ASR has been mostly focused on End-to-End non-streaming models trained on curated datasets by just ignoring labels for example wav2vec 2.0~\cite{baevski2020wav2vec}. However, we focus on demonstrating the effectiveness of in-domain SSL on uncurated, very noisy, untranscribed audio data for streaming hybrid ASR. In addition to those differences,~\cite{zhang2020pushing} focused on semi-supervised student-teacher End-to-end training while we emphasized on thorough analysis on various pre-training strategies in hybrid ASR models. From analysis perspective, our work is closest to~\cite{misra2021comparison} since they presented interesting observations for comparing supervised and unsupervised pre-training End-to-End models. However, They only focused on comparing in-domain monolingual SSL versus out-domain monolingual supervised pre-training. We extended the analysis to monolingual versus multilingual pre-training, out-domain versus in-domain pre-training, single-head versus multi-head multilingual SSL, and supervised versus unsupervised pre-training in streaming hybrid ASR with training on uncurated audio data. In addition to all, we presented fully automated efficient data pre-processing approach and shown its effectiveness.

\section{Self-supervised learning for hybrid ASR}
\label{sec:methods}

We summarize our fully automated SSL for streaming hybrid ASR in this section. Firstly, we will propose a fully automated data pre-processing pipeline to utilize the wild, uncurated audio data. Secondly, we will introduce a SSL pre-training named Lfb2vec motivated from~\cite{zhang2020pushing}. Thirdly, we will discuss original contrastive loss named InfoNCE~\cite{baevski2020wav2vec}, its potential drawbacks and recently developed alternative to overcome them~\cite{chen2021simpler}. Fourthly we will introduce the multi-head multilingual SSL model. Fifthly, we will compare pre-training strategies such as monolingual/multi-lingual supervised pre-training and out-domain SSL pre-training. Lastly, we will explain how to do fine-tuning on streaming hybrid ASR model from any of the pre-training strategies.

\subsection{Data pre-processing pipeline}
\label{ssec:preprocess}

In order to utilize the wild audio data, we propose the following pre-processing pipeline. Firstly, we convert all the audio data to 16 khz 16 bit PCM through FFMPEG. Secondly, we use Voice Activity Detection (VAD) model to filter and remove the long silences (more than 1 second). Then, based on the long silences detected by VAD, we segment all audio data to segments with max length of 20 seconds. Later, we convert audio segments to 80-dim log-Mel features, with feature processing time window as 25-ms with 10-ms window shift. Finally, we utilize Audio Event Detection (AED) model to distinguish the speech portion of the audio from background noises such as music. The AED is a Xception-small model adapted from~\cite{chollet2017xception} with several internal layers removed. It was trained as a 14-class classification problem on an internal dataset with 800,000 two-second audio examples from various sources, each of which contains one or more positive classes. The classes covered are \{'alarm', 'bark', 'clapping', 'crosstalk', crowd', 'crying', 'engine', 'explosion',
'gunshot', 'laughter', 'music', 'screaming', 'siren', and 'speech'\}. In order to utilize the AED's prediction, we consider the following two AED filters (i) \emph{Speech-filter}: ignore utterance which does not have any speech event; (ii) \emph{Speech-crop}: Crop the utterances based on speech event to only include the speech portion; and one general filter (iii) \emph{Rand-crop}: random crop for long utterances ($\ge$ 5s) which can be perceived as data augmentation. AED filters will be used on-the-fly during training.

\subsection{Pre-training: Lfb2vec}
\label{ssec:pretrain}
\begin{figure}[tb]
\begin{minipage}[b]{1.0\linewidth}
  \centering
  \centerline{\includegraphics[width=9cm]{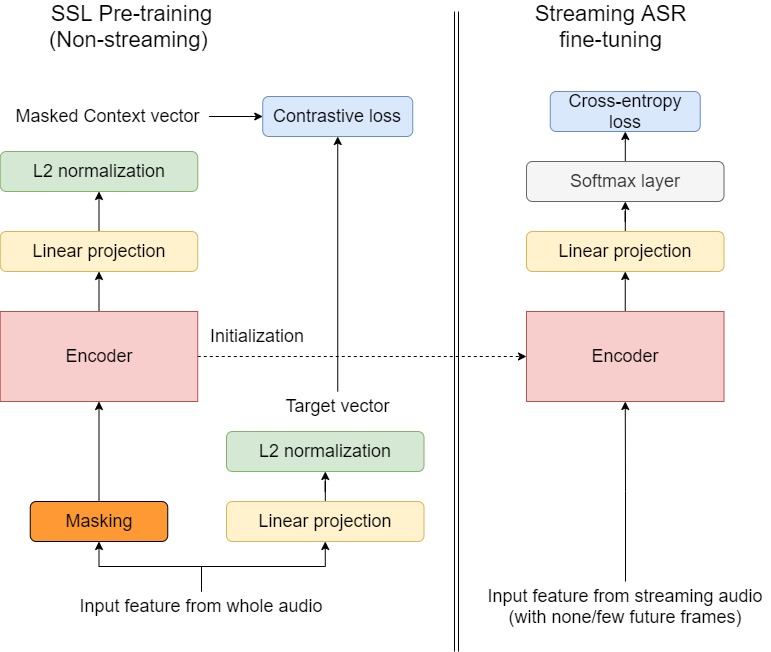}}
\end{minipage}
\caption{Overall overview of SSL pre-training through Lfb2vec.}
\label{fig:lfb2vec}
\end{figure}

Motivated from wav2vec 2.0~\cite{baevski2020wav2vec} and~\cite{zhang2020pushing}, we propose to use Lfb2vec for SSL pre-training. The overall Lfb2vec procedure is depicted on left side of figure~\ref{fig:lfb2vec}. On one side, log-Mel features are masked through identical procedure as~\cite{zhang2020pushing}. Initial time steps are sampled randomly to be masked with probability 0.065 and then we mask the subsequent 10 time steps. Later, masked features are fed to Encoder, 20-dimensional linear projection layer and L2 normalized to yield masked context vectors. In this study, we utilized 6-layers Bi-directional LSTMs~\cite{schuster1997bidirectional} with 600 hidden dimensions as our Encoder but any model such as Transformers or conformers can be used. On the other side, log-Mel features are fed to another 20-dimensional linear projection layer and L2 normalized to yield target vectors. We finally optimize the contrastive loss between masked positions of context vector and target vectors. 100 negative samples are drawn randomly from same utterance but other positions of target vectors. 

We utilize Lfb2vec for (1) in-domain monolingual SSL, (2) out-domain monolingual SSL, and (3) out-domain multilingual SSL motivated from~\cite{conneau2020unsupervised}. Since finding 100 negative samples from the same utterance is impossible for streaming SSL with short chunk length. Therefore, we train Lfb2vec in non-streaming pre-training by feeding the whole utterance's log-Mel features to the model motivated from~\cite{wang2021unispeech}.  



\subsection{Contrastive loss functions}
\label{ssec:loss}
InfoNCE is the popular contrastive loss which has been used in Computer Vision (CV)~\cite{chen2020simple,goyal2021self}, Natural Language Processing (NLP)~\cite{devlin2018bert} and wav2vec 2.0 in ASR~\cite{baevski2020wav2vec}. It has been shown that InfoNCE has connection in estimating \textit{mutual information} with low variance~\cite{chen2020simple}. InfoNCE's loss is formulized as:

\begin{equation}
    \mathrm{InfoNCE}(x,y^{1:N}) = f(x,y_1)-\log(\frac{1}{N}\sum_{j=1}^{N} e^{f(x,y_j)})
\end{equation}

Where $y_1$ is the true positive and $y^{2:N}$ are the negative samples. Also, $f(x,y)$ is Cosine similarity. However, recently a couple of InfoNCE's drawbacks have drawn attentions. Specifically, (i) InfoNCE is sensitive to mini batch size and requires large negative samples, therefore it can be computationally a burden and inefficient for small mini-batch sizes. (ii) InfoNCE's loss is bounded by $\log$(\#samples) and is biased estimate of \textit{mutual information}. To overcome these issues in CV and NLP field, a novel contrastive loss function has been introduced such as flatNCE~\cite{chen2021simpler}. Mathematical formulation of flatNCE is shown below:
\begin{equation}
    \mathrm{flatNCE}(x,y^{1:N}) =\frac{\log(\frac{1}{N}\sum_{j=1}^{N} e^{f(x,y_j)})-f(x,y_1)}{\mathrm{detach}[\frac{1}{N}\sum_{j=1}^{N} e^{f(x,y_j)})-f(x,y_1)]}
\end{equation}

where $\mathrm{detach}[f_\theta(x)]$ is an operation that bars gradient back-propagation. We can readily observe that flatNCE is a flat function where $\mathrm{flatNCE}(x,y^{1:N}) = 1 \quad \forall \, x, y^{1:N}$  which fulfills the zero-variance property. Therefore, flatNCE can be considered as a self-normalized contrastive loss. From gradient and differentiable optimization perspective flatNCE is very close to InfoNCE. Also interestingly, they have rigorously shown that flatNCE is conjugate dual of InfoNCE \cite{chen2021simpler}. Due to the space limitation, we won't provide more detailed information about these loss functions and encouraged readers to look through the provided references. 

\subsection{Multi-head multilingual self supervised learning}

Inspired from multilingual supervised pre-training introduced by~\cite{huang2013cross}, we developed a multi-head multilingual self supervised learning. Particularly, in our SSL modeling the main encoder will be shared for all languages followed by a language dependent linear projection layer. We experimented both only out-domains 8 locals multi-head SSL and combination of out-domains and in-domain 9 locals multi-head SSL.  
\subsection{Supervised pre-training}
\label{ssec:pretrain}
As an alternative to SSL pre-training through Lfb2vec, we explore monolingual or multilingual supervised pre-training approaches. As monolingual supervised pre-training, we train a 600 hidden dimension 6-layer Latency-Controlled BLSTM ( LC-BLSTM)~\cite{xue2017improving} followed by linear projection layer on English data. As multilingual supervised pre-training, motivated from~\cite{huang2013cross}, we train a shared 600 hidden dimension 6-layer LC-BLSTM~\cite{xue2017improving} followed by language dependent linear projection layer and softmax layer on 8 locales: English, Chinese, Japanese, Italian, French, Spanish, German and Brazilian Portuguese. Both monolingual and multilingual supervised pre-training are trained in streaming setting with 20 frames chunk length, 20 frames look ahead length, and cross entropy loss function.


\subsection{Fine-tuning for streaming hybrid ASR}
\label{ssec:finetune}

We fine-tune the supervised or unsupervised pre-trained model with hybrid streaming ASR model. Specifically, we use 6-layer LC-BLSTM~\cite{xue2017improving} followed by linear projection layer as our senone-based Acoustic Model (AM). We train the AM with 20 frames chunk length, 20 frames look ahead length, and cross entropy loss function on target language. We consider two-stage fine-tuning approach since the pre-training models can only initialize the LC-BLSTM layers and not the linear projection layer. Therefore, we freeze the LC-BLSTM layers initialized from pre-training model and only train the linear projection layer initialized from scratch. Then, in second stage, we train the whole model. In the decoding stage, we use a 5-gram Language Model (LM) with vocabulary of over 1M words.

\section{Experiments and results}
\label{sec:results}

\begin{table}
\centering
\caption{WERs(\%) for effect of AED filters on data pre-processing pipeline}
\smallskip
\renewcommand{\arraystretch}{1.6}
\label{tab:preprocess}       
\resizebox{0.95\columnwidth}{!}{%
\begin{tabular}{c|cc|c|cc}
\hline \hline\noalign{\smallskip}
\multicolumn{1}{c|}{} & \multicolumn{2}{c|}{AED Filtering}  & \multicolumn{1}{c|}{Other filters} & \multicolumn{2}{c}{Results}\\
 SSL  & Speech-filter  & Speech-crop & Rand-crop &  Overall & Multi-speaker  \\
    &  &  &  &   &  conversation \\
\noalign{\smallskip}\hline
 \xmark & - & - & - &10.90 & 23.75 \\
\noalign{\smallskip}\hline
 \cmark & \xmark & \xmark & \xmark &10.05 & 19.48 \\
 \cmark & \cmark & \xmark & \xmark &10.13 & 19.27 \\
 \cmark & \cmark & \cmark & \xmark &10.03 & 19.13 \\
 \cmark & \cmark & \cmark & \cmark &\textbf{9.93} & \textbf{18.99} \\
\noalign{\smallskip}\hline\hline
\end{tabular}
}
\end{table}

\begin{table*}
\centering
\caption{WERs(\%) for comparison of pre-training strategies.}
\smallskip
\renewcommand{\arraystretch}{1}
\label{tab:pretrain}       
\resizebox{2\columnwidth}{!}{%
\begin{tabular}{cccc|cp{35mm}|cccc}
\hline \hline\noalign{\smallskip}
\multicolumn{4}{c|}{Pre-training method} & \multicolumn{2}{c|}{Pre-training data} &
\multicolumn{4}{c}{Results}\\
 Supervised & SSL & loss & Streaming & Monolingual & Multilingual & Dictation& Single-speaker  &  Multi-speaker & Overall  \\
 & & & &  & & & conversation &  conversation &  \\
\noalign{\smallskip}\hline
\xmark & \xmark & \xmark & \xmark & \xmark & \xmark   &11.09 & 5.71 & 23.75 &10.90\\
\noalign{\smallskip}\hline\hline
single-head & \xmark & CE & \cmark& out-domain (en), 11K & \xmark   & 10.46 & 5.43 & 22.12 &  10.27\\
single-head & \xmark & CE & \cmark& out-domain (en), 77K & \xmark   &10.38 & 5.59 & 20.95 & 10.15 \\
multi-head & \xmark & CE & \cmark& \xmark & out-domain, 12.7K   &10.34 & 5.19 & 19.95 & 9.95 \\
\noalign{\smallskip}\hline\hline
single-head & \xmark & CE & \xmark& out-domain (en), 11K & \xmark   & 10.42&5.49  & 22.61 &  10.29\\
single-head & \xmark & CE & \xmark& out-domain (en), 77K & \xmark   &10.53 & 5.24 & 21.08 & 10.19 \\
multi-head & \xmark & CE & \xmark& \xmark & out-domain, 12.7K   &10.42 & 5.25 & 20.18 & 10.04 \\
\noalign{\smallskip}\hline\hline
\xmark & single-head & InfoNCE& \xmark& out-domain (en), 77K & \xmark  & 10.47 &  5.60 & 19.60 & 10.10\\
\xmark & single-head & InfoNCE& \xmark& \xmark & out-domain, 12.7K   &10.64 & 5.46 & 20.17 & 10.24\\
\xmark & multi-head & InfoNCE& \xmark& \xmark & out-domain, 12.7K   & 10.44& 5.56 & 19.56 & 10.07\\
\noalign{\smallskip}\hline\hline
\xmark & single-head & InfoNCE& \xmark& in-domain (ro), 11.7K & \xmark &  10.36 & \textbf{5.15} & 18.41 & \textbf{9.83}\\
\xmark & single-head & flatNCE& \xmark& in-domain (ro), 11.7K & \xmark &  \textbf{10.29} & 5.36 &\textbf{18.21} & \textbf{9.82}\\
\noalign{\smallskip}\hline\hline
\end{tabular}
}
\end{table*}

\subsection{Datasets and experimental setup}
\label{ssec:data}

We evaluate the effectiveness of SSL pre-training in hybrid ASR on our internal transcribed Romanian dataset. We used 716 hours for training set and three test sets (i) Dictation set: 6.2 hours of audio data; (ii) Single-speaker conversation: 10 hours of presentations, lectures, etc; and (iii) Multi-speaker conversation: 2.76 hours of interview. For pre-training stage, we utilize three internal datasets: (i) unlabeled Romanian audio data: we collect 11.7K hours of audio from YouTube including news, podcasts, interviews and etc; (ii) out-domain monolingual dataset: we consider 77K hours of transcribed English audio data; (iii) out-domain multilingual dataset: we collect almost 12.7K of audio data with 36.76\% English, 15.04\% Chinese, 6.68\% Japanese, 8.36\% Italian, 10.37\% French, 9.44\% Spanish, 7.15\% German and 6.20\% Brazilian Portuguese languages. Dataset (i) has been used only for SSL, But datasets (ii) and (iii) have been used for both SSL (by ignoring labels) and supervised pre-training.

We train monolingual supervised pre-training on 77K hours of English data through 8 epochs of Adam optimizer with max learning rate of 2e-4. We train multilingual supervised pre-training on 12.7K hours of 8 languages through 6 epochs of Adam optimizer with max learning rate of 1e-4. We train SSL pre-training on both 77K hours of English data and multilingual data through 15M and 10M steps of AdamW~\cite{loshchilov2017decoupled} optimizer with max learning rate of 1e-3, respectively. Similarly, we train SSL pre-training on Romanian data through either Adam or AdamW max learning rate of 1e-3 with various learning rate scheduling. In all of the SSL pre-training, we use the first 10\% of steps as warm-up to reach the max learning rate and the rest of 90\% with linear decay to reach 5e-6. In all fine-tuning stages, we train for 5 epochs through Adam optimizer with learning rate of 1.6e-3 and 8e-4 in first and second half of training.  

\subsection{Results and discussion}
\label{ssec:optimization-results}

\begin{figure}[tb]
\begin{minipage}[b]{1.0\columnwidth}
  \centering
  \centerline{\includegraphics[width=8cm]{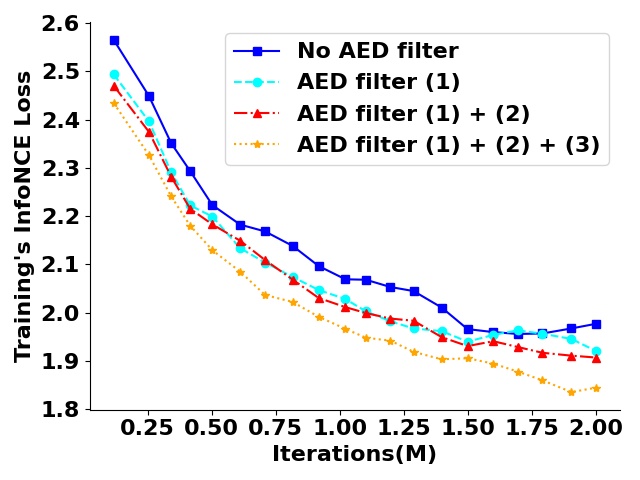}}
\end{minipage}
\caption{Effect of AED filters on training InfoNCE loss.}
\label{fig:result_loss}
\end{figure}

Word error rate (WER) is used to evaluate the performance of ASR. The unlabeled Romanian audio data, collected from YouTube, is closest to our hardest test set "Multi-speaker conversation". That is the main reason we can observe substantial gain by using the SSL on this dataset. To present the effectiveness of data pre-processing pipeline, we trained all the SSL model with 2 million steps of Adam optimizer with or without AED filters. Table~\ref{tab:preprocess} and Figure~\ref{fig:result_loss} show the effectiveness of AED filters. Based on Figure~\ref{fig:result_loss}, with the increase of the addition of the AED's filters, we can observe decrease in the InfoNCE loss such that $l(\text{no AED}) > l(\text{AED filter (1)}) > l(\text{AED filters (1) + (2)}) > l(\text{all filters})$. AED filters (1) and (1) + (2) reduce unlabeled data by 5\% and 13\%, respectively. AED filters will remove the non-speech part of audio data which makes it easier for SSL model to learn representation useful for ASR downstream task. The decrease in the loss is correlated to 2.5\% and 1.2\% decrease in WER for multi-speaker conversation and overall test set.

\begin{figure}[tb]
\begin{minipage}[b]{1.0\linewidth}
  \centering
  \centerline{\includegraphics[width=8.7cm]{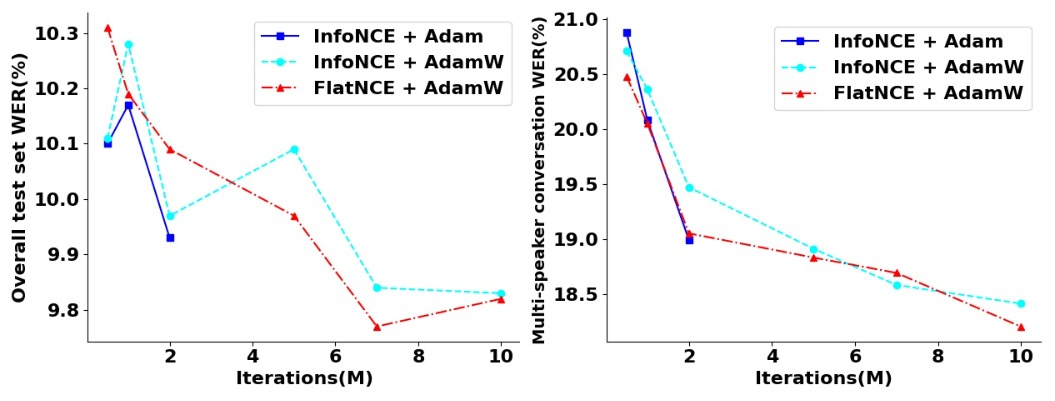}}
\end{minipage}
\caption{Effect of optimizers and learning rate scheduling for SSL. Since Adam optimizer is unstable for learning rate scheduling with larger than 2 million iterations, it has been presented only until 2 million iterations.}
\label{fig:result_optimizer}
\end{figure}

SSL training usually requires 100 sweeps to learn useful representation in comparison to supervised cross entropy training which only needs 8-10 sweeps. We have observed that, by increasing the number of iterations and keeping the same high learning rate (usually 1e-3), Adam optimizer will be very unstable. Figure~\ref{fig:result_optimizer} show the comparison of loss functions and optimizers over different learning rate scheduling (different maximum number of iterations, number of steps for warm-up). We can observe that InfoNCE with Adam optimizer is performing well in learning rate scheduling with low number of iterations but will be unstable for learning rate scheduling with larger number of iterations by keeping the same high learning rate. However, AdamW~\cite{loshchilov2017decoupled} can overcome the instability issue in Adam for both InfoNCE and flatNCE. Moreover, we can observe that WER will decrease for both overall and multi-speaker conversation test sets by increase of maximum number of steps in learning rate scheduling. Also, generally flatNCE has 1\%-2\% lower WER in comparison to InfoNCE in most of the experiments. 

We present a thorough comparison of different pre-training strategies in Table~\ref{tab:pretrain}. Generally we can observe that InfoNCE (flatNCE) SSL with in-domain pre-training can reduce the WER by 6.5\% (7.2\%), 9\% (6.1\%), 22.4\% (23.3\%), 9.8\% (9.8\%) for InfoNCE (flatNCE) in dictation, single-speaker conversation, multi-speaker conversation and overall test sets in comparison from training from scratch, respectively. flatNCE can slightly outperform InfoNCE. Most importantly, in-domain flatNCE SSL pre-training can outperform the best alternative pre-training strategy, supervised multilingual pre-training, by 8.7\% and  1.3\% in multi-speaker conversation and overall test sets, respectively. As expected the biggest gain comes from the multi-speaker conversation set which is the closest to the unlabeled YouTube data. 

Now, we will discuss about multilingual versus monolingual pre-training. Based on Table~\ref{tab:pretrain}, multilingual streaming supervised pre-training can outperform monolingual streaming supervised pre-training by 9.8\%(4.7\%) and 3.1\%(1.9\%) in multi-speaker conversation and overall sets with almost equal (6 times less) data, respectively. Also, in non-streaming setting, multilingual non-streaming supervised pre-training can outperform monolingual non-streaming supervised pre-training by 10.7\%(4.2\%) and 2.4\%(1.4\%) in multi-speaker conversation and overall sets with almost equal (6 times less) data, respectively. However, it is vice versa for SSL pre-training if we use single-head for both scenarios. The monolingual SSL pre-training can outperform single-head multilingual SSL pre-training by 2.8\% and 1.3\% in multi-speaker conversation and overall sets, respectively. It seems that learning general and robust representation based on multilingual data through~\cite{conneau2020unsupervised} is harder than monolingual case and it might not be achieved just by using all the data simultaneously without any distinguishing between the languages. Maybe as it needs more innovations such as using language-dependent softmax layer which has been used in supervised pre-training~\cite{huang2013cross}. As expected with multi-head multilingual SSL can be on-par or a little better than monolingual SSL with 6 times less data.

Now we will discuss about streaming versus non-streaming supervised pre-training. Streaming supervised pre-training can outperform non-streaming supervised pre-training for both monolingual  and multilingual settings. Specifically, streaming multilingual supervised pre-training can outperform non-steaming multilingual supervised pre-training by 1\% in both multi-speaker conversation and overall sets. 

Now, we will discuss about streaming supervised pre-training versus non-streaming SSL pre-training. Streaming supervised pre-training is learning the representation based on the out-domain senone labels and also is matched with the streaming fine-tuning stage. On the other side, non-streaming SSL pre-training is learning the representation based on the audio data itself by constructing positive and negative samples. However, since we need to collect 100 negative samples for every frame and in streaming training we are limited to 40 time frames, then it is not possible to perform streaming SSL in the current setting. In this case, the SSL pre-training and following  fine-tuning are essentially mismatched in terms of streaming and non-streaming. However, this seems not a severe problem. The mismatched SSL pre-training is still very effective in boosting the accuracy of the final model. Based on the results presented in Table~\ref{tab:pretrain}, taking the monolingual out-domain pre-training data as an example, non-streaming SSL can get roughly on-par results as the  supervised streaming pre-training overall, 10.10 vs 10.15. When using 11.7k in-domain data in SSL, it outperforms any supervised pre-training significantly on all test sets.

\section{CONCLUSIONS}
\label{sec:conclusions}
Thorough comparison of pre-training strategies for hybrid streaming ASR shows that self-supervised learning can reduce Word error rate of in-domain test set by 7\% in comparison to all the alternative pre-training strategies with out-domain data including all the combinations of monolingual or multilingual versus supervised or self-supervised pre-training strategies. In order to fully utilizing the power of self-supervised learning for uncurated, wild audio data collected from YouTube we show that (i) Xception-based accurate Audio Event Detection (AED) model in data pre-processing pipeline can boost the performance by 2.5\% in in-domain test set; (ii) Adam optimizer is unstable in learning rate scheduling with large maximum iterations while AdamW can overcome the instability and boost the performance more than 3\%; (iii) InfoNCE contrastive loss functions has large biased estimation of mutual information and newly developed flatNCE contrastive loss in NLP and CV domain can overcome it and boost the performance up to 2\%; (iv) multi-head multilingual SSL can outperform single-head multilingual SSL by 1.6\%.




\bibliographystyle{IEEEtran}

\bibliography{mybib}

\end{document}